# A feasible algorithm for typing in Elementary Affine Logic

Patrick Baillot
Laboratoire d'Informatique de Paris-Nord /CNRS
Université Paris-Nord, France
pb@lipn.univ-paris13.fr

Kazushige Terui
National Institute of Informatics
Tokyo, Japan
terui@nii.ac.jp

**Abstract**

We give a new type inference algorithm for typing lambda-terms in Elementary Affine Logic (EAL), which is motivated by applications to complexity and optimal reduction. Following previous references on this topic, the variant of EAL type system we consider (denoted $EAL^{\star}$) is a variant without sharing and without polymorphism. Our algorithm improves over the ones already known in that it offers a better complexity bound: if a simple type derivation for the term $t$ is given our algorithm performs $EAL^{\star}$ type inference in polynomial time.

## 1 Introduction

Since [GSS92, Gir98], Linear logic (LL) has been shown a fruitful logical setting in which computational complexity can be brought into the picture of the proofs-as-programs correspondence. In particular Light linear logic ([Gir98]) and Soft linear logic ([Laf04]) are variants of LL in which all numerical functions programmed are polynomial time. Another system, Elementary linear logic (ELL, see [Gir98, DJ03]) corresponds to Kalmar elementary complexity.

Hence one can consider specific term calculi designed through the Curry-Howard correspondence and program directly in these languages with the guaranteed complexity bound ([Rov98, Ter01]). However this turns out in practice to be a difficult task, in particular because these languages require managing specific constructs corresponding to the logical modalities. Considering the *affine* variant (i.e. with unrestricted weakening) of these systems is an advantage ([Asp98]) but does not suppress the difficulty.

An alternative point of view is to keep ordinary lambda-calculus and use the logic as a type system: then if a program is well-typed the logic provides a way to execute it with the guaranteed complexity bound. The difficulty is then moved to the problem of type inference.

This approach and the corresponding type inference problems have been studied in [CM01, CRdR03] for Elementary affine logic (EAL) and [Bai02, Bai04] for Light affine logic (LAL). It was shown that type inference in the propositional fragments of these systems is decidable. Typing in EAL is actually also motivated by another goal (see [CM01, ACM00]): EAL terms can be evaluated with the optimal reduction discipline much more easily than general terms, by using the abstract part of Lamping's algorithm. Thus EAL typing can be seen as an intermediate step which, if it succeeds, allows to speed up optimal reduction.

However though these type inference problems have been shown decidable the algorithms provided, either for EAL or LAL, are not really efficient. They all run at least in exponential time, even if one considers as input a simply typed lambda-term. Our goal is to improve this state-of-the-art by providing more efficient and possibly more simple algorithms. Our motivation is typing in Dual light affine logic (DLAL, [BT04]) which is a simplification of LAL and corresponds to Ptime, but here as a first step we propose a new procedure for EAL.

**Contribution.** Technically speaking the difficulty with EAL typing is to find out *where* in the derivation to place !-rules and *how many* of them. This corresponds in proof-nets terminology to placing *boxes*. The algorithms in [CM01] and [CRdR03] are based on two tactics for *first* placing abstract boxes and *then* working out their number using linear constraints. Our approach also uses linear constraints but departs from this point of view by determining the place of boxes *dynamically*, at the time of constraints solving. This method was actually already proposed in [Bai02] for LAL typing but with several conditions; in particular the term had to be in normal form. In the present work we show that in a system without sharing (like DLAL, but unlike LAL) this approach is considerably simplified. In particular it results that:

- one can use as intermediary syntax a very simple term calculus (introduced in [AR02]) instead of proof-nets like in [Bai02];
- the procedure can be run in polynomial time, if one considers as input a simply typed lambda-term (instead of an untyped lambda-term).

**Outline.** The paper will proceed as follows: in section 2 we introduce Elementary affine logic and the type system $EAL^\star$ we consider for lambda-calculus; in section 3 we describe the term calculus (*pseudo-terms*, or *concrete syntax*) we will use to denote $EAL^\star$ derivations and we prove a theorem (Theorem 7) on $EAL^\star$ typability; finally in section 4 we give an $EAL^\star$ decoration algorithm (based on Theorem 7), prove it can be run in polynomial time (4.2) and derive from it an $EAL^\star$ type inference algorithm (4.3).

**Notations.** Given a lambda-term $M$ we denote by $FV(M)$ the set of its free variables. Given a variable $x$ we denote by $no(x, M)$ the number of occurrences of $x$ in $M$. We denote by $|M|$ the structural size of a term $M$. The notation $\longrightarrow$ will stand for $\beta$-reduction on lambda-terms. We denote substitution (without capture of variable) by $M[N/x]$. When there is no ambiguity we will write $M[M_i/x_i]$ for $M[M_1/x_1, \dots, M_n/x_n]$.

Notations for lists: $\epsilon$ will denote the empty list and pushing element $a$ on list $l$ will be denoted by $a :: l$. The prefix relation on lists will be denoted by $\leq$.

## 2 Typing in Elementary Affine Logic

The formulas of Intuitionistic multiplicative Elementary affine logic ( Elementary affine logic for short, EAL) are given by the following grammar:

$$A, B ::= \alpha \mid A \multimap B \mid !A \mid \forall \alpha . A$$

We restrict here to propositional EAL (without quantification). A natural deduction presentation for this system is given on Figure 1.

$$\frac{}{A \vdash A}\ \text{(var)} \qquad \frac{\Gamma \vdash B}{\Gamma, A \vdash B}\ \text{(weak)}$$

$$\frac{\Gamma_1 \vdash A \multimap B \quad \Gamma_2 \vdash A}{\Gamma_1, \Gamma_2 \vdash B}\ \text{(appl)} \qquad \frac{\Gamma, A \vdash B}{\Gamma \vdash A \multimap B}\ \text{(abst)}$$

$$\frac{\Gamma_1 \vdash !A_1 \cdots \Gamma_n \vdash !A_n \quad A_1, \ldots, A_n \vdash B}{\Gamma_1, \ldots, \Gamma_n \vdash !B}\ \text{(prom)}$$

$$\frac{\Gamma \vdash !A \quad !A, \ldots, !A, \Delta \vdash B}{\Gamma, \Delta \vdash B}\ \text{(contr)}$$

Figure 1: Natural deduction for EAL.

We call *erasure* $A^-$ of an EAL formula $A$ the simple type defined inductively by:

$$\alpha^- = \alpha,\ (!A)^- = A^-,\ (A \multimap B)^- = A^- \to B^-.$$

Conversely, given a simple type $T$ we say that an EAL formula $A$ is a *decoration* of $T$ if we have $A^- = T$.

We will use EAL as a type system for lambda-terms, but in a way more constrained than that allowed by this natural deduction presentation:

**Definition 1** *Let $M$ be a lambda-term; we say $M$ is typable in $EAL^\star$ with type $\Gamma \vdash M : A$ if there is a derivation of this judgment in the system from Figure 3 where any (contr) rule is either followed by an (abst) on the contracted formula, or is the last rule of the derivation.*

The main restriction of this definition is that it does not allow *sharing* for typing lambda-terms. This comes in contrast with the computational study of ELL carried out for instance in [DJ03] but is motivated by several points:

- if we use sharing for typing, some important structure of the proof is lost when we look at the term after forgetting the type derivation and the computational behaviour of the proof and the term will differ (see the discussion in [BT04]);

$$\dfrac{\dfrac{x_1 : !A, \ldots, x_n : !A, \Gamma \vdash M : B}{x : !A, \Gamma \vdash M[x/x_1, \ldots, x_n] : B}\ \text{(contr)}}{x : !!A, !\Gamma \vdash M[x/x_1, \ldots, x_n] : !B}\ \text{(prom)}$$

$$\Downarrow$$

$$\dfrac{\dfrac{x_1 : !A, \ldots, x_n : !A, \Gamma \vdash M : B}{x_1 : !!A, \ldots, x_n : !!A, \Gamma \vdash M : !B}\ \text{(prom)}}{x : !!A, !\Gamma \vdash M[x/x_1, \ldots, x_n] : !B}\ \text{(contr)}$$

Figure 2:

- this approach without sharing is enough to define Dual Light Affine Logic (DLAL) typing ([BT04]) which is sufficient to capture polynomial time computation;
- sharing-free derivations are necessary to be able to use EAL for optimal reduction with the abstract part of Lamping's algorithm, as argued by Coppola and Martini in [CM01];
- finally: using sharing would make type inference more difficult ...

Actually, the notion of EAL-typability for lambda-terms considered by Coppola and Martini in [CM01] even if it does not allow sharing is slightly more liberal than $EAL^\star$-typability. In their case one might follow a (contr) rule by a (prom) rule provided all the $M_i$s are variables (we can then keep only the main premise of the rule). However it follows from [CRdR03] (with the notion of canonical abstract terms) that if a judgment $\Gamma \vdash M : A$ is derivable in Coppola-Martini's type system, then it is derivable in $EAL^\star$. The key point for this remark is the fact that one can perform the commutation of Figure 2. In proof-net words ([DJ03]) this means pushing the contraction nodes which are premises of an auxiliary door of a box outside the box.

$$\dfrac{}{x : A \vdash x : A}\ \text{(var)} \qquad \dfrac{\Gamma \vdash M : B}{\Gamma, x : A \vdash M : B}\ \text{(weak)}$$

$$\dfrac{\Gamma_1 \vdash M_1 : A \multimap B \quad \Gamma_2 \vdash M_2 : A}{\Gamma_1, \Gamma_2 \vdash (M_1 M_2) : B}\ \text{(appl)} \qquad \dfrac{\Gamma, x : A \vdash M : B}{\Gamma \vdash \lambda x.M : A \multimap B}\ \text{(abst)}$$

$$\dfrac{\Gamma_1 \vdash M_1 : !A_1 \cdots \Gamma_n \vdash M_n : !A_n \quad x_1 : A_1, \ldots, x_n : A_n \vdash M : B}{\Gamma_1, \ldots, \Gamma_n \vdash M[M_i/x_i] : !B}\ \text{(prom)}$$

$$\dfrac{x_1 : !A, \ldots, x_n : !A, \Delta \vdash M : B}{x : !A, \Delta \vdash M[x/x_1, \ldots, x_n] : B}\ \text{(contr)}$$

Figure 3: Typing rules for $EAL^\star$.

# 3 Concrete syntax and box reconstruction

## 3.1 Pseudo-terms

In order to describe the structure of type derivations we need a term calculus more informative than lambda-calculus. We will use the language introduced in [AR02] (called *concrete syntax* in this paper), which is convenient because it has no explicit construct neither for boxes, nor for contractions. It was stressed in this reference that this syntax is not faithful for LAL: several type derivations (LAL proofs) correspond to the same term. However it is faithful for $EAL^\star$, precisely because there is no sharing and no ambiguity on the placement of contractions.

Let us introduce *pseudo-terms*:

$$t, u ::= x \mid \lambda x.t \mid (t)u \mid !t \mid \bar{!}t$$

The idea is that ! constructs correspond to main doors of boxes in *proof-nets* ([Gir87, AR02]) while $\bar{!}$ constructs correspond to auxiliary doors of boxes. But note that there is no information in the pseudo-terms to link occurrences of ! and $\bar{!}$ corresponding to the same box.

There is a natural erasure map $(.)^-$ from pseudo-terms to lambda-terms consisting in removing all occurrences of ! and $\bar{!}$. When $t^- = M$, $t$ is called a *decoration* of $M$.

For typing pseudo-terms the rules are the same as in Definition 1 and Figure 3, but for (prom):

$$\frac{\Gamma_1 \vdash t_1 : !A_1 \cdots \Gamma_n \vdash t_n : !A_n \quad x_1 : A_1, \ldots, x_n : A_n \vdash t : B}{\Gamma_1, \ldots, \Gamma_n \vdash !t\ [\bar{!}t_i/x_i] : !B} \text{ (prom)}$$

We want to give an algorithm to determine if a pseudo-term can be typed in $EAL^\star$: this can be seen as a kind of correctness criterion allowing to establish if boxes can be reconstructed in a suitable way; this issue will be examined in 3.2.

Actually, when searching for $EAL^\star$ type derivations for (ordinary) lambda-terms it will be interesting to consider a certain subclass of derivations. A type derivation in $EAL^\star$ is *restricted* if in all applications of the rule (prom),

(i) the subject $M$ of the main premise $x_1 : A_1, \ldots, x_n : A_n \vdash M : B$ is not a variable, and

(ii) the last rules to derive auxiliary premises $\Gamma_i \vdash M_i :!A_i$ ($1 \leq i \leq n$) are either (var) or (appl).

A pseudo-term is *restricted* if it is obtained by the following grammar:

$$\begin{array}{rcl} a & ::= & x \mid \lambda x.t \mid (t)t \\ t & ::= & !^m a, \end{array}$$

where $m$ is an arbitrary value in $\mathbb{Z}$ and $!^m a$ is defined by:

$$\begin{array}{rcll} !^m a & = & \underbrace{! \cdots !}_{m\ times}\ a & \text{if } m \geq 0; \\ & = & \underbrace{\bar{!} \cdots \bar{!}}_{-m\ times}\ a & \text{if } m < 0. \end{array}$$

We then have:

**Proposition 1**

1. *(For lambda-terms) if $\Gamma \vdash M : A$ has a type derivation, then it also has a restricted type derivation.*

2. *(For pseudo-terms) Every restricted derivation yields a restricted pseudo-term.*

*Proof (Sketch).* 1. Notice that the typing rules in Figure 3 satisfy the following substitution property:

> if $\Gamma_1 \vdash M_1 : A$ has a derivation of length $l_1$ and $x : A, \Gamma_2 \vdash M_2 : B$ has a derivation of length $l_2$ such that no $(cntr)$ rule has been performed on $x : A$, then $\Gamma_1, \Gamma_2 \vdash M_2[M_1/x] : B$ has a derivation of length shorter than $l_1 + l_2$.

Given a derivation of $\Gamma \vdash M : A$ that is not restricted and contains for instance

$$\dfrac{\begin{array}{c}\vdots\\ \Gamma_1 \vdash N_1 :!A_1\end{array} \quad \dfrac{}{y : A_1 \vdash y : A_1}\ (\text{var})}{\Gamma_1 \vdash N_1 :!A_1}\ (\text{prom})$$

violating the condition (i), one can rewrite it into

$$\begin{array}{c}\vdots\\ \Gamma_1 \vdash N_1 :!A_1,\end{array}$$

strictly shortening the length of the derivation.

Given a derivation that is not restricted and contains

$$\dfrac{\dfrac{\begin{array}{c}\vdots\\ \Gamma_1 \vdash N :!C\end{array} \quad \begin{array}{c}\vdots\\ y : C \vdash M_1 : A_1\end{array}}{\Gamma_1 \vdash M_1[N/y] :!A_1}\ (\text{prom}) \qquad \begin{array}{c}\vdots\\ x_1 : A_1 \vdash M : B\end{array}}{\Gamma_1 \vdash M[M_1[N/y]/x_1] :!B}\ (\text{prom})$$

violating the condition (ii), we have $y : C \vdash M[M_1/x_1] : B$ by the substitution property. Therefore, one can rewrite the derivation into

$$\dfrac{\begin{array}{c}\vdots\\ \Gamma_1 \vdash N :!C\end{array} \quad \begin{array}{c}\vdots\\ y : C \vdash M[M_1/x_1] : B\end{array}}{\Gamma_1 \vdash M[M_1/x_1][N/y] :!B}\ (\text{prom}),$$

strictly shortening the length of the derivation.

The proof is similar when $M$ contains more than one free variables. The other cases are immediate.

2. By induction on the length of the restricted derivation.

### 3.2 Box reconstruction

We will consider words over the language $\mathcal{L} = \{!, \bar{!}\}^\star$.

If $t$ is a pseudo-term and $x$ is an occurrence of variable (either free or bound) in $t$, we define $t\langle x\rangle$ as the word of $\mathcal{L}$ obtained by listing the occurrences of !, $\bar{!}$ holding $x$ in their scope. More formally:

$$
\begin{array}{rcl}
x\langle x\rangle & = & \epsilon, \\
(t_1)t_2\langle x\rangle & = & t_i\langle x\rangle \text{ where } t_i \text{ is the subterm containing } x, \\
(\lambda y.t)\langle x\rangle & = & t\langle x\rangle \ (y \text{ might be equal to } x), \\
(!t)\langle x\rangle & = & ! :: (t\langle x\rangle), \\
(\bar{!}t)\langle x\rangle & = & \bar{!} :: (t\langle x\rangle).
\end{array}
$$

We define a map: $s : \mathcal{L} \to \mathbb{Z}$ by:

$$
\begin{array}{rcl}
s(\epsilon) & = & 0, \\
s(! :: l) & = & 1 + s(l) \\
s(\bar{!} :: l) & = & -1 + s(l)
\end{array}
$$

We call $s(l)$ the *sum* associated to $l$.

Let $t$ be a pseudo-term. We say that $t$ satisfies the *bracketing condition* if:

- for any occurrence of variable $x$ in $t$,

$$\forall l \leq t\langle x\rangle,\ s(l) \geq 0,$$

- moreover if $x$ is an occurrence of free variable:

$$s(t\langle x\rangle) = 0.$$

That is to say: if ! is seen as an opening bracket and $\bar{!}$ as a closing bracket, in $t\langle x\rangle$ any $\bar{!}$ matches a ! (we will say that $t\langle x\rangle$ is *weakly well-bracketed*) and if $x$ is free $t\langle x\rangle$ is well-bracketed.

We say $t$ satisfies the *scope condition* if: for any subterm $\lambda x.v$ of $t$, for any occurrence $x_i$ of $x$ in $v$, $v\langle x_i\rangle$ is well-bracketed:

- $\forall l \leq v\langle x_i\rangle, \quad s(l) \geq 0$,
- and $s(v\langle x_i\rangle) = 0$.

It is obvious that:

**Lemma 2** *If $t$ is a pseudo-term which satisfies the scope condition, then any subterm of $t$ also satisfies this condition.*

**Proposition 3** *If $t$ is an $EAL^\star$ typed term, then $t$ satisfies the bracketing and scope conditions.*

*Proof.* By induction on the $EAL^\star$ type derivations.

Now, we can observe the following property:

**Lemma 4 (Boxing)** *If $!u$ is a pseudo-term which satisfies the bracketing condition then there exist $v$, $u_1, \ldots, u_n$ unique (up to renaming of $v$'s free variables) such that:*

- *$FV(v) = \{x_1, \ldots, x_n\}$ and for $1 \leq i \leq n$, $no(x_i, v) = 1$,*
- *$!u = !v[\bar{!}u_1/x_1, \ldots, \bar{!}u_n/x_n]$,*
- *for $1 \leq i \leq n$, $v\langle x_i \rangle$ is well-bracketed.*

*Proof.* We denote by $!_0$ the first occurrence of $!$ in the term considered: $!_0 u$. Denote by $\bar{!}_1, \ldots, \bar{!}_n$ the occurrences of $\bar{!}$ matching $!_0$ in the words $!u\langle x \rangle$, where $x$ ranges over the occurrences of variables in $!u$. Let $u_i$, with $1 \leq i \leq n$, be the subterms of $!u$ such that $\bar{!}_i u_i$ is a subterm of $!u$, for $1 \leq i \leq n$. Then it is clear that no $u_i$ is a subterm of a $u_j$, for $i \neq j$. Let now $v$ be the pseudo-term obtained from $u$ by replacing each $\bar{!}_i u_i$ by a distinct variable $x_i$. Then naturally we have $!u = !v[\bar{!}u_1/x_1, \ldots, \bar{!}u_n/x_n]$, and by definition of $\bar{!}_i$ we know that for $1 \leq i \leq n$, $v\langle x_i \rangle$ is well-bracketed.

Finally let us assume $x$ is an occurrence of free variable in $v$ distinct from $x_i$, for $1 \leq i \leq n$. Then $x$ is an occurrence of free variable in $!u$, and as $!u$ is well-bracketed we have that $s(!u\langle x \rangle) = 0$, hence $x$ is in the scope of a $\bar{!}_0$ matching $!_0$. Then $\bar{!}_0$ must be one of the $\bar{!}_i$, for $1 \leq i \leq n$, hence $x$ is in $u_i$ and thus does not occur in $v$, which gives a contradiction. Therefore we have $FV(v) = \{x_1, \ldots, x_n\}$ and the proof is over.

Given a pseudo-term $t$ we call *EAL type assignment* for $t$ a map $\Gamma$ from the variables of $t$ (free or bound) to $EAL$ formulas. EAL type assignments are simply called assignments when there is no danger of confusion. This map $\Gamma$ is extended to a partial map from subterms of $t$ to $EAL$ formulas by the following inductive definition:

$$\begin{array}{lcll}
\Gamma(!u) & = & !A, & \text{if } \Gamma(u) = A, \\
\Gamma(\bar{!}u) & = & A, & \text{if } \Gamma(u) = !A, \text{ undefined otherwise}, \\
\Gamma(\lambda x.u) & = & A \multimap B, & \text{if } \Gamma(x) = A, \Gamma(u) = B, \\
\Gamma((u_1)u_2) & = & B, & \text{if } \Gamma(u_2) = A \text{ and } \Gamma(u_1) = A \multimap B, \text{ undefined otherwise.}
\end{array}$$

Given a pair $(t, \Gamma)$ of a pseudo-term $t$ and an assignment $\Gamma$ (we omit $\Gamma$ if it is natural from the context) we say that $(t, \Gamma)$ satisfies the *typing condition* if:

- $\Gamma(t)$ is defined (so in particular each subterm of $t$ of the form $(u_1)u_2$ satisfies the condition above),
- for any variable $x$ of $t$ which has at least 2 occurrences we have: $\Gamma(x)$ is of the form $!B$ for some formula $B$.

Given an $EAL^\star$ type derivation for a pseudo-term $t$ there is a natural assignment $\Gamma$ obtained from this derivation: the value of $\Gamma$ on free variables is obtained from the environment of the final judgment and its value on bound variables from the type of the variable in the premise of the abstraction rule in the derivation.

**Proposition 5** *If $t$ is an $EAL^\star$ typed pseudo-term and $\Gamma$ is an associated assignment then $(t, \Gamma)$ satisfies the typing condition.*

Moreover it is easy to observe that:

**Lemma 6** *If $(t, \Gamma)$ satisfies the typing condition and $u$ is a subterm of $t$, then $(u, \Gamma)$ also satisfies the typing condition.*

Now, the conditions on pseudo-terms we have listed up to now are sufficient to ensure that $t$ is an $EAL^{\star}$ typed pseudo-term:

**Theorem 7** *If $t$ is a pseudo-term and $\Gamma$ an assignment such that:*

- *$t$ satisfies the bracketing and scope conditions,*
- *$(t, \Gamma)$ satisfies the typing condition,*

*then $t$ is typable in $EAL^{\star}$ with a judgment $\Delta \vdash t : A$ such that: $\Gamma(t) = A$ and $\Delta$ is the restriction of $\Gamma$ to the free variables of $t$.*

*Proof.* Let us use the following numeration for the conditions:
(i) bracketing, (ii) scope, (iii) typing.
We proceed by induction on the pseudo-term $t$:

- $t = x$ is trivial.
- $t = \lambda x.u$,

  it is clear that $u$ satisfies the first part of the bracketing condition. The second part of the bracketing condition (for free variables) is ensured by the fact that $t$ satisfies the scope condition for $x$. It is then trivial that $u$ satisfies conditions (ii), (iii), thus by induction hypothesis we have in $EAL^{\star}$ : $\Delta, x : A \vdash u : B$ where $\Gamma(x) = A$, $\Gamma(u) = B$ and by an abstraction rule we get the expected property for $t$.

- $t = (t_1)t_2$,

  the subterms $t_1$, $t_2$ then satisfy conditions (i) to (iii), hence by induction hypothesis we have:

  $$\Delta_1 \vdash t_1 : A_1$$
  $$\Delta_2 \vdash t_2 : A_2.$$

  where $\Gamma(t_i) = A_i$ and $\Delta_i$ is the restriction of $\Gamma$ to the free variables of $t_i$. As $t$ satisfies the typing condition (iv) we know that $A_1$ is of the form $A_1 = A_2 \multimap B_1$. If $t_1$ and $t_2$ have a free variable $y$ in common then as $t$ satisfies the typing condition we have that $\Gamma(y) = !B$. We rename in $t_1$, $t_2$ the free variables that they have in common, and from the previous judgments applying an (appl) rule and a (contr) rule we get the expected judgment for $t$.

- $t = \bar{!}u$,

  then $t$ does not satisfy the bracketing condition (i), so the implication is valid.

- $t = !u$,

  by the Boxing Lemma 4, $t$ can be written as $t = !v[\bar{!}u_1/x_1, \dots, \bar{!}u_n/x_n]$ where $FV(v) = \{x_1, \dots, x_n\}$ and each $v\langle x_i\rangle$ is well-bracketed.

  Let us show that $u_i$ satisfies conditions (i)–(iii). Take $y$ an occurrence of variable in $u_i$. We have:

  $$t\langle y\rangle = ! :: v\langle x_i\rangle :: \bar{!} :: u_i\langle y\rangle,$$

  thus as $v\langle x_i\rangle$ is well bracketed, $u_i\langle y\rangle$ satisfies the bracketing condition and $u_i$ satisfies (i).

  By Lemmas 2 and 6 as $t$ satisfies (ii) and (iii), $u_i$ also satisfies (ii) and (iii). Therefore by induction hypothesis we get that there exists an $EAL^\star$ derivation of conclusion:

  $$\Delta_i \vdash u_i : A_i,$$

  where $A_i = \Gamma(u_i)$, for $1 \leq i \leq n$.

  Let us now examine the conditions for $v$. As $t$ satisfies the bracketing condition and by the Boxing Lemma 4, we get that $v$ satisfies (i). By the Boxing Lemma again we know that all free variables of $v$ have exactly one occurrence. It is easy to check that as $t$ satisfies the scope condition (ii), so does $v$.

  Consider now the typing condition. Let $\tilde{\Gamma}$ be defined as $\Gamma$ but $\tilde{\Gamma}(x_i) = \Gamma(\bar{!}u_i)$ for $1 \leq i \leq n$. If $y$ has several occurrences in $v$ then it has several occurrences in $t$, hence $\Gamma(y) = !B$, so $\tilde{\Gamma}(y) = !B$. If $(v_1)v_2$ is a subterm of $v$ then $(v_1')v_2'$, where $v_i' = v_i[\bar{!}u_1/x_1, \dots, \bar{!}u_n/x_n]$, is a subterm of $t$ and $\tilde{\Gamma}(v_i') = \Gamma(v_i)$. Therefore as $(t, \Gamma)$ satisfies the typing condition, then so does $(v, \tilde{\Gamma})$.

  As $\Gamma(u_i) = A_i$ and $\Gamma(\bar{!}u_i)$ is defined we have $A_i = !B_i$ and $\tilde{\Gamma}(x_i) = B_i$. Finally as $v$ satisfies conditions (i)–(iii), by i.h. there exists an $EAL^\star$ derivation of conclusion:

  $$\Delta, x_1 : B_1, \dots, x_n : B_n \vdash v : C,$$

  where $C = \tilde{\Gamma}(v)$.

  If $u_i$ and $u_j$ for $i \neq j$ have a free variable $y$ in common then as $t$ satisfies the typing condition we have $\Gamma(y) = !B$. We rename the free variables common to several of the $u_i$s, apply a (prom) rule to the judgements on $u_i$ and the judgement on $v$, then some (contr) rules and get a judgement:

  $$\Delta' \vdash t : !C.$$

  Hence the i.h. is valid for $t$, which concludes the proof.

# 4 A decoration algorithm

## 4.1 Decorations and instantiations

We consider the following *decoration problem*:

**Problem 1 (decoration)** *let $x_1 : A_1, \ldots, x_n : A_n \vdash M : B$ be a simply typed term; does there exist EAL decorations $A'_i$ of the $A_i$ for $1 \leq i \leq n$ and $B'$ of $B$ such that $x_1 : A'_1, \ldots, x_n : A'_n \vdash M : B'$ is a valid $EAL^\star$ judgement for $M$?*

For that we will need to find out the possible concrete terms corresponding to $M$. Actually following section 3.1 and Prop. 1 it is sufficient to search for a suitable term in the set of restricted pseudo-terms, instead of considering the whole set of pseudo-terms. To perform this search we will use *parameterized restricted pseudo-terms* (parameterized pseudo-terms, for short), defined by the following grammar:

$$\begin{array}{rcl} a & ::= & x \mid \lambda x.t \mid (t)t \\ t & ::= & !^{\mathbf{n}} a \end{array}$$

where $\mathbf{n}$ is a fresh parameter (meant to range over $\mathbb{Z}$).

Given a parameterized pseudo-term we denote by $par(t)$ the set of its parameters. An *instantiation* $\phi : par(t) \rightarrow \mathbb{Z}$ allows to define a restricted pseudo-term $\phi(t)$ obtained by substituting each parameter $\mathbf{n}$ by the integer $\phi(\mathbf{n})$.

We will also consider *parameterized types* defined by:

$$A ::= !^{\mathbf{n}}\alpha \mid !^{\mathbf{n}}(A \multimap A)$$

where $\mathbf{n}$ is a fresh parameter.

We denote by $par(A)$ the set of parameters of $A$. If $\phi$ is an instantiation $\phi : par(A) \rightarrow \mathbb{Z}$, then $\phi(A)$ is defined only when a nonnegative integer is substituted for each parameter. We define the size $|A|$ of a parameterized formula $A$ as the structural size of its underlying simple type (so the sum of the number of $\multimap$ connectives and atomic subtypes).

Just as we have defined EAL type assignments for pseudo-terms we will consider *parameterized type assignments* for parameterized pseudo-terms with values parameterized types, and *simple type assignments* for lambda-terms with values simple types. Let $\Sigma$ be a parameterized type assignment for a parameterized pseudo-term $t$. We denote by $par(\Sigma)$ the set of parameters occurring in parameterized types $\Sigma(x)$, for all variables $x$ of $t$. Let $\phi : par(\Sigma) \rightarrow \mathbb{Z}$ be an instantiation and suppose that $\phi(\mathbf{n}) \geq 0$ holds for every $\mathbf{n} \in par(\Sigma)$. Then one can define the map $\phi\Sigma$ by: $\phi\Sigma(x) = \phi(\Sigma(x))$. When it is defined, it is an EAL type assignment for $\phi(t)$. We define the size $|\Sigma|$ of $\Sigma$ as the maximum of $|\Sigma(x)|$ for all variables $x$.

The erasure map $(.)^-$ is defined for parameterized pseudo-terms and parameterized types analogously to those for pseudo-terms and EAL types. It is clear that given a lambda-term $M$ there exists a unique parameterized pseudo-term $t$ (up to renaming of its parameters) such that $t^- = M$. We denote $t$ by $\overline{M}$ and call it the *parameter decoration* of $M$. Note that the size of $\overline{M}$ is linear in the size of $M$. Given a simple type $T$, its *parameter decoration* $\overline{T}$ is defined analogously. Finally, given a simple type assignment $\Theta$ for a lambda-term $t$ (with values simple types), its *parameter decoration* $\overline{\Theta}$ is defined pointwise, by taking $\overline{\Theta}(x) = \overline{\Theta(x)}$, where all these decorations are taken with disjoint parameters.

The following picture illustrates the relationship among various notions introduced so far:

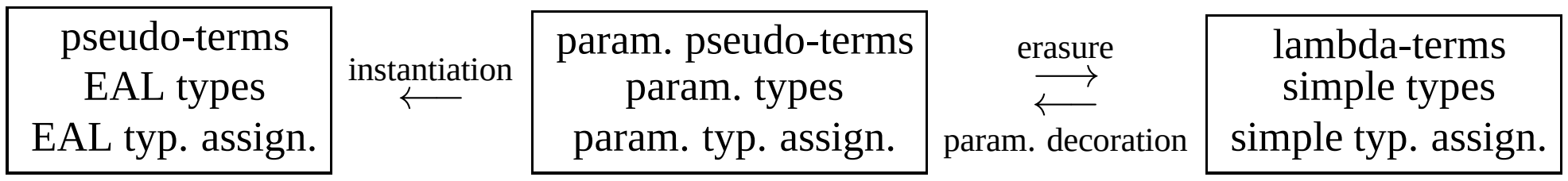

Given a simple type derivation of $x_1 : T_1, \ldots, x_n : T_n \vdash M : T$, one can naturally obtain a simple type assignment $\Theta$ for $M$. Furthermore, it is automatic to build parameter decorations $\overline{M}$ and $\overline{\Theta}$. Suppose now that there is an instantiation $\phi$ for $(\overline{M}, \overline{\Theta})$ for which $\phi\overline{\Theta}$ is defined. Then $\phi\overline{\Theta}(x_i)$ is a decoration of $T_i$ for $1 \leq i \leq n$ and $\phi\overline{\Theta}(\overline{M})$ is a decoration of $T$. Conversely, any decorations of $T_i$'s and $T$ are obtained through some instantiations for $(\overline{M}, \overline{\Theta})$. Therefore, the decoration problem boils down to the following *instantiation problem*:

**Problem 2** *Given a parameterized pseudo-term $t$ and a parameterized type assignment $\Sigma$ for it: does there exist an instantiation $\phi$ such that $\phi(t)$ has an $EAL^\star$ type derivation associated to $\phi\Sigma$?*

To solve this problem we will use Theorem 7 to find suitable instantiations $\phi$ if there exists any. For that we will need to be able to state the conditions of this theorem on parameterized pseudo-terms; they will yield linear constraints. We will speak of *linear inequations*, meaning in fact both linear equations and linear inequations.

We will consider lists over parameters $\mathbf{n}$. Let us denote by $\mathcal{L}'$ the set of such lists.

As for pseudo-terms we define for $t$ a parameterized pseudo-term and $x$ an occurrence of variable in $t$, a list $t\langle x\rangle$ in $\mathcal{L}'$ by:

$$\begin{aligned} x\langle x\rangle &= \epsilon, \\ (t_1)t_2\langle x\rangle &= t_i\langle x\rangle \text{ where } t_i \text{ is the subterm containing } x, \\ (\lambda y.t)\langle x\rangle &= t\langle x\rangle (y \text{ might be equal to } x), \\ (!^{\mathbf{n}}a)\langle x\rangle &= \mathbf{n} :: (a\langle x\rangle). \end{aligned}$$

The sum $s(l)$ of an element $l$ of $\mathcal{L}'$ is a linear combination defined by:

$$\begin{aligned} s(\epsilon) &= 0, \\ s(\mathbf{n} :: l) &= \mathbf{n} + s(l). \end{aligned}$$

Let $t$ be a parameterized pseudo-term. We define the *boxing constraints* for $t$ as the set of linear inequations $\mathcal{C}^b(t)$ obtained from $t$ in the following way:

- bracketing: for any occurrence of variable $x$ in $t$ and any prefix $l$ of $t\langle x\rangle$, add the inequation: $s(l) \geq 0$; moreover if $x$ is an occurrence of free variable add the equation $s(t\langle x\rangle) = 0$.
- scope: for any subterm $\lambda x.v$ of $t$, for any occurrence $x_i$ of $x$ in $v$, add similarly the inequations expressing the fact that $v\langle x_i\rangle$ is well-bracketed.

It is then straightforward that:

**Proposition 8** *Given an instantiation $\phi$ for $t$, we have: $\phi(t)$ satisfies the bracketing and scope conditions iff $\phi$ is a solution of $\mathcal{C}^b(t)$.*

Note that the number of inequations in $\mathcal{C}^b(t)$ is polynomial in the size of $t$ (hence also in the size of $t^-$).

In the sequel, we will need to unify parameterized types. For that, given 2 parameterized types $A$ and $B$ we define their unification constraints $U(A,B)$ by:

$$\begin{array}{rcl} U(!^{\mathbf{m}}\alpha, !^{\mathbf{n}}\alpha) & = & \{\mathbf{m}=\mathbf{n}\} \\ U(!^{\mathbf{m}}(A_1 \multimap A_2), !^{\mathbf{n}}(B_1 \multimap B_2)) & = & \{\mathbf{m}=\mathbf{n}\} \cup U(A_1,B_1) \cup U(A_2,B_2) \end{array}$$

and $U(A,B)=\{false\}$ (unsolvable constraint) in the other cases.

Let $\Sigma$ be a parameterized type assignment for a parameterized pseudo-term $t$. Then we extend $\Sigma$ to a partial map from the subterms of $t$ to parameterized types in the following way:

$$\begin{array}{lcll} \Sigma(!^{\mathbf{n}}a) & = & !^{\mathbf{m}}A \text{ with } \mathbf{m} \text{ fresh}, & \text{if } \Sigma(a) = !^{\mathbf{k}}A, \\ \Sigma(\lambda x.u) & = & !^{\mathbf{m}}(A \multimap B) \text{ with } \mathbf{m} \text{ fresh}, & \text{if } \Sigma(x)=A, \Sigma(u)=B, \\ \Sigma((u_1)u_2) & = & B, & \text{if } \Sigma(u_1) = !^{\mathbf{n}}(A \multimap B), \text{ undefined otherwise.} \end{array}$$

We define the *typing constraints* for $(t,\Sigma)$ as the set of linear inequations $\mathcal{C}^{typ}(t,\Sigma)$ obtained from $t$, $\Sigma$ as follows:

- abstractions: for any subterm of $t$ of the form $\lambda x.u$ with $\Sigma(\lambda x.u) = !^{\mathbf{m}}(A \multimap B)$, add $\mathbf{m}=0$.
- applications: for any subterm of $t$ of the form $(u_1)u_2$ with $\Sigma(u_1) = !^{\mathbf{m}}(A_1 \multimap B_1)$ and $\Sigma(u_2)=A_2$ add the constraints $U(A_1,A_2)\cup\{\mathbf{m}=0\}$; if $\Sigma(u_1)$ is not of this form then $\mathcal{C}^{typ}(t,\Sigma)=\{false\}$.
- bang: for any subterm of $t$ of the form $!^{\mathbf{n}}u$ with $\Sigma(!^{\mathbf{n}}u) = !^{\mathbf{m}}A$ and $\Sigma(u) = !^{\mathbf{k}}A$, add the constraints $\mathbf{m}=\mathbf{k}+\mathbf{n}$ and $\mathbf{m}\geq 0$.
- contractions: for any variable $x$ of $t$ which has at least 2 occurrences and $\Sigma(x) = !^{\mathbf{m}}A$, add the constraint $\mathbf{m}\geq 1$.
- types: for any parameter $\mathbf{m}$ in $par(\Sigma)$, add the constraint $\mathbf{m}\geq 0$.

We then have:

**Proposition 9** *Let $t$ be a parameterized pseudo-term and $\Sigma$ be a parameterized type assignment for $t$ such that $\Sigma(t)$ is defined. Given an instantiation $\phi$ for $(t,\Sigma)$, we have: $\phi\Sigma$ is defined and $(\phi(t),\phi\Sigma)$ satisfies the typing condition iff $\phi$ is a solution of $\mathcal{C}^{typ}(t,\Sigma)$.*

Note that the number of inequations in $\mathcal{C}^{typ}(t,\Sigma)$ is polynomial in $(|t|+|\Sigma|)$.

We define $\mathcal{C}(t,\Sigma)=\mathcal{C}^b(t)\cup\mathcal{C}^{typ}(t,\Sigma)$. Using the two previous Propositions and Theorem 7 we get the following result, which solves the instantiation problem:

**Theorem 10** *Let $t$ be a parameterized pseudo-term, $\Sigma$ be a parameterized type assignment for $t$ such that $\Sigma(t)$ is defined, and $\phi$ be an instantiation for $t$, $\Sigma$. The two following conditions are equivalent:*

- *$\phi(t)$ is typable in $EAL^\star$ with a judgment $\Delta \vdash \phi(t) : A$ such that $\phi\Sigma(t) = A$ and $\Delta$ is the restriction of $\phi\Sigma$ to the free variables of $t$,*

- *$\phi$ is a solution of $\mathcal{C}(t, \Sigma)$.*

*Moreover the number of inequations in $\mathcal{C}(t, \Sigma)$ is polynomial in $(|t| + |\Sigma|)$.*

If $t$ and $\Sigma$ come from a simply typed lambda-term $M$ and its typing derivation, then $\Sigma(t)$ is always defined and $\mathcal{C}^{typ}(t, \Sigma)$ never gives rise to $false$. By noting this fact, we obtain the following result, which solves the decoration problem:

**Theorem 11** *Let $x_1 : A_1, \ldots, x_n : A_n \vdash M : B$ be a simply typed term and let $\Theta$ be the associated simple type assignment. There exist decorations $A'_i$ of the $A_i$ for $1 \leq i \leq n$ and $B'$ of $B$ such that $x_1 : A'_1, \ldots, x_n : A'_n \vdash M : B'$ is a valid $EAL^\star$ judgement iff there is a solution $\phi$ to $\mathcal{C}(\overline{M}, \overline{\Theta})$.*

*In this case each solution $\phi$ gives a suitable $EAL^\star$ judgment $x_1 : A'_1, \ldots, x_n : A'_n \vdash M : B'$. Moreover the number of inequations and the number of parameters in $\mathcal{C}(\overline{t}, \overline{\Theta})$ are polynomial in $(|t| + |\Theta|)$.*

We give an example of execution of the algorithm in the Appendix.

## 4.2 Solving the constraints

Now we turn our attention to the constraints and their solutions. Let $t$ be a parameterized pseudo-term and $\Sigma$ be an assignment. We consider instead of the previous instantiation maps with values in $\mathbb{Z}$, maps with rational numbers as values: $\psi : par(t) \cup par(\Sigma) \rightarrow \mathbb{Q}$.

If $\psi$ is such a map and $a$ is a non-negative integer we defined the map $a\psi$ by: $(a\psi)(\mathbf{n}) = a.\psi(\mathbf{n})$, for any parameter $\mathbf{n}$.

**Lemma 12** *If $\psi$ is a solution of $\mathcal{C}(t, \Sigma)$ and $a$ is a strictly positive integer then $a\psi$ is also a solution of $\mathcal{C}(t, \Sigma)$.*

*Proof.* It is enough to observe that for any inequation of $\mathcal{C}^b(t)$ and $\mathcal{C}^{typ}(t, \Sigma)$ if $\psi$ is a solution then so is $a\psi$:

- all inequations from $\mathcal{C}^b(t)$ and all those from $\mathcal{C}^{typ}(t, \Sigma)$ except the contractions case are homogeneous (no constant element in combinations) and as $a \geq 0$ the inequalities are preserved when multiplying both members by $a$;

- the inequations coming from the contraction cases in $\mathcal{C}^{typ}(t, \Sigma)$ are of the form $\mathbf{m} \geq 1$, so as $a \geq 1$ we have: if $\psi(\mathbf{m}) \geq 1$ holds then so does $a\psi(\mathbf{m}) \geq 1$.

Recall that the problem of finding if a linear system of inequations $\mathcal{C}$ admits a solution in $\mathbb{Q}$ can be solved in polynomial time in the size of $\mathcal{C}$ and its number of variables. Hence we have:

**Proposition 13** *The problem of whether the system $\mathcal{C}^{typ}(t, \Sigma)$ admits a solution with values in $\mathbb{Z}$ can be solved in time polynomial in $(|t| + |\Sigma|)$.*

*Proof.* As the number of inequations and the number of parameters in $\mathcal{C}^{typ}(t,\Sigma)$ is polynomial in $(|t|+|\Sigma|)$ and by the result we recalled above we have: one can decide if $\mathcal{C}^{typ}(t,\Sigma)$ admits a solution with values in $\mathbb{Q}$ in time polynomial in $(|t|+|\Sigma|)$.

Then, if there is no solution in $\mathbb{Q}$ there is no solution in $\mathbb{Z}$. Otherwise if $\psi$ is a solution in $\mathbb{Q}$ take for $a$ the least multiple of the denominators of $\psi(n)$, for all parameters $n$. Then by Lemma 12, $a\psi$ is a solution in $\mathbb{Z}$. It then follows that:

**Theorem 14** *The decoration problem of Theorem 11 can be solved in time polynomial in* $(|t|+|\Gamma|)$.

## 4.3 Type inference

The procedure for $EAL^\star$ decoration we have given can be extended to a type inference procedure for $EAL^\star$ in the way used in [CM01]: given an ordinary term $M$,

- compute the principal assignment $\Theta$ for $M$ (giving the principal simple type),
- use the procedure of Theorem 11 to find if $M$, $\Theta$ admits a suitable $EAL^\star$ decoration.

It follows from a result of [CRdR03] that:

**Proposition 15** *if $M$ is $EAL^\star$ typable and admits as principal simple type judgment $\Delta \vdash M : A$, then $M$ admits an $EAL^\star$ type judgment which is a decoration of this judgment.*

In order to have a self-contained presentation and to take advantage of the simplicity of our framework we will give a proof of Prop. 15 here. It follows from it that the algorithm for $EAL^\star$ type inference we gave is complete.

First we define two functions $T_E(.)$ and $\mathcal{E}(.)$ on pseudo-terms allowing to find for a pseudo-term $t$ all its possible $EAL^\star$ types: $T_E(.)$ gives a typing scheme and $\mathcal{E}(.)$ the associated set of equations. Note that the term $t$ might not be $EAL^\star$ typable anyway as we are not considering here the boxing conditions. The functions are defined by induction on pseudo-terms below:

- if $t = x$:
  then $T_E(t) =< x : \alpha; \alpha >$, $\mathcal{E}(x) = \emptyset$.
- if $t = \lambda x.t_1$ and $T_E(t_1) =< \Gamma; B >$:
  then $T_E(t) =< \Gamma'; A \multimap B >$ with: $A = \Gamma(x)$ and $\Gamma' = \Gamma$ if $\Gamma(x)$ is defined; $A = \alpha$ (fresh variable) and $\Gamma'$ extends $\Gamma$ with $\Gamma'(x) = \alpha$ otherwise. $\mathcal{E}(t) = \mathcal{E}(t_1)$.
- if $t = (t_1)t_2$ and $T_E(t_i) =< \Gamma_i; A_i >$ for $i = 1, 2$:
  let $FV(t_1) \cap FV(t_2) = \{x_1, \ldots, x_k\}$, $\Gamma$ be defined by: $\Gamma(y) = \Gamma_i(y)$ if $y \in FV(t_i)$ and $y \notin \{x_1, \ldots, x_k\}$; $\Gamma(x_j) = !\beta_j$ for $1 \leq j \leq k$, where the $\beta_j$s are fresh type variables. Let $T_E(t) =< \Gamma; \alpha >$ ($\alpha$ fresh variable) and $\mathcal{E}(t) = \mathcal{E}(t_1) \cup \mathcal{E}(t_2) \cup \{A_1 \equiv (A_2 \multimap \alpha); \Gamma_1(x_j) \equiv !\beta_j, \Gamma_2(x_j) \equiv !\beta_j, 1 \leq j \leq k\}$.

- if $t = !t_1$ and $T_E(t_1) =< \Gamma_1; A_1 >$ then $T_E(t) =< \Gamma_1; !A_1 >$ and $\mathcal{E}(t) = \mathcal{E}(t_1)$.
- if $t = \bar{!}t_1$ and $T_E(t_1) =< \Gamma_1; A_1 >$ then $T_E(t) =< \Gamma_1; \alpha >$ ($\alpha$ fresh variable) and $\mathcal{E}(t) = \mathcal{E}(t_1) \cup \{A_1 \equiv !\alpha\}$.

Call *EAL substitution* (resp. *simple type substitution*) a map $\sigma$ from type variables to EAL formulas (resp. simple types). Given an EAL-substitution $\sigma$ and an EAL formula $A$, $\sigma A$ is the formula obtained by substituting type variables $\alpha$ in $A$ by $\sigma\alpha$. Given a set of equations $\mathcal{E}$ we say that $\sigma$ is *a solution of* $\mathcal{E}$ if for any $A_1 \equiv A_2$ in $\mathcal{E}$, $\sigma A_1 = \sigma A_2$ holds.

We have:

**Proposition 16** *Let $t$ be a pseudo-term. The two following conditions are equivalent:*

- $(t, \Gamma)$ *satisfies the typing condition and* $\Gamma(t) = B$*;*
- $T_E(t) =< \Delta, A >$ *and there exists a solution $\sigma$ of $\mathcal{E}(t)$ such that:* $\Gamma = \sigma\Delta$ *and* $B = \sigma A$*.*

Now, we define similar functions $T_S(.)$ and $\mathcal{S}(.)$ for typing terms in simple types:

- if $t = x$:
  then $T_S(t) =< x : \alpha; \alpha >$, $\mathcal{S}(x) = \emptyset$.
- if $t = \lambda x.t_1$ and $T_S(t_1) =< \Gamma; B >$:
  then $T_S(t) =< \Gamma'; A \multimap B >$ with: $A = \Gamma(x)$ and $\Gamma' = \Gamma$ if $\Gamma(x)$ is defined; $A = \alpha$ (fresh variable) and $\Gamma'$ extends $\Gamma$ with $\Gamma'(x) = \alpha$ otherwise. $\mathcal{S}(t) = \mathcal{S}(t_1)$.
- if $t = (t_1)t_2$ and $T_S(t_i) =< \Gamma_i; A_i >$ for $i = 1, 2$:
  let $FV(t_1) \cap FV(t_2) = \{x_1, \dots, x_k\}$, $\Gamma$ be defined by: $\Gamma(y) = \Gamma_i(y)$ if $y \in FV(t_i)$ and $y \notin \{x_1, \dots, x_k\}$; $\Gamma(x_j) = \beta_j$ for $1 \leq j \leq k$, where the $\beta_j$s are fresh type variables. Let $T_S(t) =< \Gamma; \alpha >$ ($\alpha$ fresh variable) and $\mathcal{S}(t) = \mathcal{S}(t_1) \cup \mathcal{S}(t_2) \cup \{A_1 \equiv (A_2 \multimap \alpha); \Gamma_1(x_j) \equiv \beta_j, \Gamma_2(x_j) \equiv \beta_j, 1 \leq j \leq k\}$.
- if $t = !t_1$ and $T_S(t_1) =< \Gamma_1; A_1 >$ then $T_S(t) =< \Gamma_1; A_1 >$ and $\mathcal{S}(t) = \mathcal{S}(t_1)$.
- if $t = \bar{!}t_1$ and $T_S(t_1) =< \Gamma_1; A_1 >$ then $T_S(t) =< \Gamma_1; \alpha >$ ($\alpha$ fresh variable) and $\mathcal{S}(t) = \mathcal{S}(t_1) \cup \{A_1 \equiv \alpha\}$.

We have:

**Proposition 17** *Let $M$ be a lambda-term and $t$ a pseudo-term such that $t^- = M$. Then $M$ has a simple type iff $\mathcal{S}(t)$ has a solution and in that case : if $\tau$ is the most general unifier (m.g.u.) of $\mathcal{S}(t)$ and $T_S(t) =< \Gamma; A >$ then $\tau\Gamma' \vdash M : \tau A$ is the principal simple type of $M$ (where $\Gamma'$ is the restriction of $\Gamma$ to $FV(M)$).*

We need to relate equations in EAL and in simple types. Let $\mathcal{E}$ be a set of EAL equations and $\mathcal{E}^-$ denote the set of equations $A_1^- \equiv A_2^-$, for all equations $A_1 \equiv A_2$ in $\mathcal{E}$.

Let $\sigma$ be an EAL substitution and $\sigma^-$ be the simple type substitution given by: $\sigma^-(\alpha) = \sigma(\alpha)^-$, for all $\alpha$. Observe that:

**Fact.** If $\sigma$ is a solution of $\mathcal{E}$ then $\sigma^-$ is a solution of $\mathcal{E}^-$.

Finally we have:

**Proposition 18** *Let $\mathcal{E}$ be a set of EAL equations. If $\mathcal{E}$ admits a solution and $\tau$ is the m.g.u. of $\mathcal{E}^-$ then there exists a solution $\sigma$ of $\mathcal{E}$ such that $\sigma^- = \tau$.*

*Proof.* It can be adapted in a straightforward way from the proof of Proposition 21 in [Bai04]. Moreover we have:

**Proposition 19** *Let $t$ be a pseudo-term and $T_E(t) =< \Gamma, A >$, $T_S(t) =< \Delta, B >$, then we have $\Delta = \Gamma^-$, $B = A^-$ and $\mathcal{S}(t) = \mathcal{E}(t)^-$.*

We can now prove Prop. 15:

*Proof.* [Prop. 15] Assume $M$ is $EAL^\star$ typable. Then there exists a pseudo-term $t$ such that $t^- = M$ and which is $EAL^\star$ typable. By Prop. 16 we know that $\mathcal{E}(t)$ admits a solution $\sigma_0$. By Prop. 19, $\mathcal{E}(t)^- = \mathcal{S}(t)$, so by the Fact observed above $\mathcal{S}(t)$ has a solution, hence it has a m.g.u. $\tau$. By Prop. 18 we get that there exists a solution $\sigma$ of $\mathcal{E}(t)$ such that $\sigma^- = \tau$.

Let $T_E(t) =< \Gamma; A >$; then by Prop. 16 and Theorem 7 we have an $EAL^\star$ judgement $\sigma\Gamma' \vdash t : \sigma A$, where $\Gamma'$ is the restriction of $\Gamma$ to $FV(t)$. Finally by Prop. 17 and 19 we know that it is a decoration of the principal simple type of $M$, which ends the proof.

It then follows from Theorem 14 that our $EAL^\star$ type inference algorithm applied to a term $M$ can be executed in time bounded by a polynomial in $(|t| + |\Theta|)$ where $\Theta$ is the principal (simple type) assignment of $M$.

Note that this does not mean that the algorithm is polynomial time in $|t|$, as it is known that the principal simple type assignment for $t$ can have a size exponential in $|t|$.

# 5 Conclusion

We have given a new type inference algorithm for $EAL^\star$ which is more efficient and we think simpler than the previous ones. It generates a set of constraints which consists of two parts: one which deals with placing suitable (potential) boxes and the other one with typing the boxed term obtained. We believe this second part could be adapted to deal with other type systems like second-order EAL (assuming a system F type given). We are currently working on the adaptation to DLAL.

We have shown that the set of constraints needed in our algorithm is polynomial in the size of the term and its simple type assignment. Finally we have also shown that by using resolution of linear inequations over rationals our algorithm can be executed in polynomial time with respect to the size of the initial term and its principal simple type assignment.

# References

[ACM00] A. Asperti, P. Coppola, and S. Martini. (Optimal) duplication is not elementary recursive. In *Proceedings POPL*, 2000.

[AR02] A. Asperti and L. Roversi. Intuitionistic light affine logic. *ACM Transactions on Computational Logic*, 3(1):1–39, 2002.

[Asp98] Andrea Asperti. Light affine logic. In *Proceedings LICS'98*. IEEE Computer Society, 1998.

[Bai02] P. Baillot. Checking polynomial time complexity with types. In *Proceedings of IFIP TCS'02*, Montreal, 2002. Kluwer Academic Press.

[Bai04] P. Baillot. Type inference for light affine logic via constraints on words. *Theoretical Computer Science*, 2004. to appear.

[BT04] P. Baillot and K. Terui. Light types for polynomial time computation in lambda-calculus. In *Proceedings of LICS'04*. IEEE Computer Press, 2004. long version on http://arXiv.org cs.LO/0402059.

[CM01] P. Coppola and S. Martini. Typing lambda-terms in elementary logic with linear constraints. In *Proceedings TLCA'01*, volume 2044 of *LNCS*, 2001.

[CRdR03] P. Coppola and S. Ronchi della Rocca. Principal typing in Elementary Affine Logic. In *Proceedings TLCA'03*, LNCS, 2003.

[DJ03] V. Danos and J.-B. Joinet. Linear logic and elementary time. Information and Computation, 2003.

[DJS94] V. Danos, J.-B. Joinet, and H. Schellinx. On the linear decoration of intuitionistic derivations. *Archive for Mathematical Logic*, 33(6), 1994.

[Gir87] J.-Y. Girard. Linear logic. *Theoretical Computer Science*, 50:1–102, 1987.

[Gir98] J.-Y. Girard. Light linear logic. *Information and Computation*, 143:175–204, 1998.

[GSS92] J.-Y. Girard, A. Scedrov, and P. Scott. Bounded linear logic: A modular approach to polynomial time computability. *Theoretical Computer Science*, 97:1–66, 1992.

[Laf04] Y. Lafont. Soft linear logic and polynomial time. *Theoretical Computer Science*, 318(1–2):163–180, 2004.

[Rov98] L. Roversi. A Polymorphic Language which is Typable and Poly-step. In *Proceedings of the Asian Computing Science Conference (ASIAN'98)*, volume 1538 of *LNCS*, pages 43 – 60. Springer Verlag, December 1998.

[Ter01] K. Terui. Light Affine Lambda-calculus and polytime strong normalization. In *Proceedings LICS'01*. IEEE Computer Society, 2001. Full version available at http://research.nii.ac.jp/∼ terui.

APPENDIX

# A An example

Let us consider a small example to illustrate our method: take $M = \lambda y.\lambda z.(y)(y)z$ (the Church integer 2). The decoration $\overline{M}$ is given by:

$$\overline{M} = !^{\mathbf{m_1}}\lambda y.!^{\mathbf{m_2}}\lambda z.!^{\mathbf{m_3}}[\,(!^{\mathbf{m_4}}y_1)\ !^{\mathbf{m_5}}[\,(!^{\mathbf{m_6}}y_2)!^{\mathbf{m_7}}z\,]\,]$$

(we have distinguished the 2 occurrences of $y$ in $y_1$ and $y_2$)

We get for the boxing constraints:

$$\mathcal{C}^b(\overline{M}) = \left\{\begin{array}{cccc}
\mathbf{m_1} & \geq & 0 & (1)\\
\mathbf{m_1}+\mathbf{m_2} & \geq & 0 & (2)\\
\mathbf{m_1}+\mathbf{m_2}+\mathbf{m_3} & \geq & 0 & (3)\\
\mathbf{m_1}+\mathbf{m_2}+\mathbf{m_3}+\mathbf{m_4} & \geq & 0 & (4)\\
\mathbf{m_1}+\mathbf{m_2}+\mathbf{m_3}+\mathbf{m_5} & \geq & 0 & (5)\\
\mathbf{m_1}+\mathbf{m_2}+\mathbf{m_3}+\mathbf{m_5}+\mathbf{m_6} & \geq & 0 & (6)\\
\mathbf{m_1}+\mathbf{m_2}+\mathbf{m_3}+\mathbf{m_5}+\mathbf{m_7} & \geq & 0 & (7)\\
\mathbf{m_2} & \geq & 0 & (8)\\
\mathbf{m_2}+\mathbf{m_3} & \geq & 0 & (9)\\
\mathbf{m_2}+\mathbf{m_3}+\mathbf{m_4} & = & 0 & (10)\\
\mathbf{m_2}+\mathbf{m_3}+\mathbf{m_5} & \geq & 0 & (11)\\
\mathbf{m_2}+\mathbf{m_3}+\mathbf{m_5}+\mathbf{m_6} & = & 0 & (12)\\
\mathbf{m_3} & \geq & 0 & (13)\\
\mathbf{m_3}+\mathbf{m_5} & \geq & 0 & (14)\\
\mathbf{m_3}+\mathbf{m_5}+\mathbf{m_7} & = & 0 & (15)
\end{array}\right.$$

where (1)–(7) express bracketing, (8)–(12) scope for $\lambda y$ and (13)–(15) scope for $\lambda z$.

Now, note that (2)–(7), (9) and (11) are consequences from the rest. Thus, $\mathcal{C}^b(\overline{M})$ is equivalent to

$$\left\{\begin{array}{cccc}
\mathbf{m_1} & \geq & 0 & (1)\\
\mathbf{m_2} & \geq & 0 & (8)\\
\mathbf{m_2}+\mathbf{m_3}+\mathbf{m_4} & = & 0 & (10)\\
\mathbf{m_2}+\mathbf{m_3}+\mathbf{m_5}+\mathbf{m_6} & = & 0 & (12)\\
\mathbf{m_3} & \geq & 0 & (13)\\
\mathbf{m_3}+\mathbf{m_5} & \geq & 0 & (14)\\
\mathbf{m_3}+\mathbf{m_5}+\mathbf{m_7} & = & 0 & (15)
\end{array}\right.$$

Now let us examine the typing constraints. We consider the principal typing assignment: $\Gamma(y) = \alpha \to \alpha$, $\Gamma(z) = \alpha$, which yields $\Gamma(M) = (\alpha \to \alpha) \to (\alpha \to \alpha)$.

Thus we have:

$\overline{\Gamma}(y) = !^{\mathbf{p_1}}(!^{\mathbf{p_2}}\alpha \multimap !^{\mathbf{p_3}}\alpha)$, $\overline{\Gamma}(z) = !^{\mathbf{p_4}}\alpha$.

We get for instance:

$$\begin{array}{lcl}
\overline{\Gamma}(!^{\mathbf{m_6}}y_2) & = & !^{\mathbf{p_6}}(!^{\mathbf{p_2}}\alpha \multimap !^{\mathbf{p_3}}\alpha)\\
\overline{\Gamma}((!^{\mathbf{m_4}}y_1)\ !^{\mathbf{m_5}}[\,(!^{\mathbf{m_6}}y_2)!^{\mathbf{m_7}}z\,]) & = & !^{\mathbf{p_3}}\alpha\\
\overline{\Gamma}(t) & = & !^{\mathbf{p_{11}}}(!^{\mathbf{p_1}}(!^{\mathbf{p_2}}\alpha \multimap !^{\mathbf{p_3}}\alpha) \multimap !^{\mathbf{p_{10}}}(!^{\mathbf{p_4}}\alpha \multimap !^{\mathbf{p_9}}\alpha))
\end{array}$$

We obtain the following typing conditions (omitting some obvious constraints):

$$
\mathcal{C}^{typ}(\overline{M}) = \left\{
\begin{array}{cclcll}
\mathbf{p_5} & = & \mathbf{m_7} + \mathbf{p_4} & \geq & 0 & (16) \\
\mathbf{p_6} & = & \mathbf{m_6} + \mathbf{p_1} & \geq & 0 & (17) \\
\mathbf{p_6} & = & 0 & & & (18) \\
\mathbf{p_2} & = & \mathbf{p_5} & & & (19) \\
\mathbf{p_7} & = & \mathbf{m_5} + \mathbf{p_3} & \geq & 0 & (20) \\
\mathbf{p_8} & = & \mathbf{m_4} + \mathbf{p_1} & \geq & 0 & (21) \\
\mathbf{p_8} & = & 0 & & & (22) \\
\mathbf{p_2} & = & \mathbf{p_7} & & & (23) \\
\mathbf{p_9} & = & \mathbf{m_3} + \mathbf{p_3} & \geq & 0 & (24) \\
\mathbf{p_{10}} & = & \mathbf{m_2} & \geq & 0 & (25) \\
\mathbf{p_{11}} & = & \mathbf{m_1} & \geq & 0 & (26) \\
\mathbf{p_1}, \ldots, \mathbf{p_4} & \geq & 0 & & & (27) \\
\mathbf{p_1} & \geq & 1 & & & (28)
\end{array}
\right.
$$

that is equivalent to:

$$
\left\{
\begin{array}{cccll}
\mathbf{p_1} & = & -\mathbf{m_6} & \geq & 1 \\
\mathbf{p_1} & = & -\mathbf{m_4} & & \\
\mathbf{p_2} & = & \mathbf{p_4} + \mathbf{m_7} & \geq & 0 \\
\mathbf{p_2} & = & \mathbf{p_3} + \mathbf{m_5} & & \\
\mathbf{p_9} & = & \mathbf{p_3} + \mathbf{m_3} & & \\
\mathbf{p_{10}} & = & \mathbf{m_2} & \geq & 0 \\
\mathbf{p_{11}} & = & \mathbf{m_1} & \geq & 0 \\
\mathbf{p_3}, \mathbf{p_4} & \geq & 0 & &
\end{array}
\right.
$$

Putting $\mathcal{C}^b(\overline{M})$ and $\mathcal{C}^{typ}(\overline{M})$ together we get that $\mathcal{C}(\overline{M})$ is equivalent to:

$$\{\mathbf{m_1}, \mathbf{m_2}, \mathbf{m_3} \geq 0; \mathbf{m_2} + \mathbf{m_3} = \mathbf{p_1} \geq 1; \mathbf{m_3} + \mathbf{m_7} = 0; \mathbf{m_5} = 0;$$

$$\mathbf{m_4} = \mathbf{m_6} = -\mathbf{p_1}; \mathbf{p_2} = \mathbf{p_3}; \mathbf{p_4} = \mathbf{p_9} = \mathbf{p_2} + \mathbf{m_3}\}$$

This finally give the following (inforamlly written) parameterized term and type with constraints, which describe all solutions to this decoration problem:

$$
\left\{
\begin{array}{c}
\overline{M} = !^{\mathbf{m_1}} \lambda y.!^{\mathbf{m_2}} \lambda z.!^{\mathbf{m_3}} [\,(\bar{!}^{\mathbf{m_2}+\mathbf{m_3}} y_1)\,[\,(\bar{!}^{\mathbf{m_2}+\mathbf{m_3}} y_2) \bar{!}^{\mathbf{m_3}} z\,]\,] \\
!^{\mathbf{m_1}}(!^{\mathbf{m_2}+\mathbf{m_3}}(!^{\mathbf{p_2}}\alpha \multimap !^{\mathbf{p_2}}\alpha) \multimap !^{\mathbf{m_2}}(!^{\mathbf{p_2}+\mathbf{m_3}}\alpha \multimap !^{\mathbf{p_2}+\mathbf{m_3}}\alpha)) \\
\text{constraints: } \{\mathbf{m_1}, \mathbf{m_2}, \mathbf{m_3}, \mathbf{p_2} \geq 0, \mathbf{m_2} + \mathbf{m_3} \geq 1\}.
\end{array}
\right.
$$

Observe that this representation corresponds to several canonical forms (6 in this particular example) in the approach of Coppola and Ronchi della Rocca (see [CRdR03]).